# Physical and Structural Design of Fast Extraction Kickers for CSNS/RCS

WANG Lei(王磊)[1] KANG Wen(康文)[1] HAO Yao-Dou(郝耀斗)[1] CHEN Yuan(陈沅)[1]
HUO Li-Hua(霍丽华)[1]
1. Institute of High Energy Physics, Chinese Academy of Sciences, Beijing 100049, China

**Abstract**：China Spallation Neutron Source (CSNS) is a high intensity beam facility being built now in China. Three kicker assemblies, eight pulsed magnets, will be used in the CSNS rapid circle synchrotron (RCS). The physical and structural designs of eight kicker magnets that are grouped in 5 different types are presented. The results of OPERA-3D simulation show that magnet center field integral meet the physics requirements of design by choosing a suitable magnet coil structure. Field uniformity for 60% width is ±0.7%. The ferrite magnet structure and composition is introduced, and the high voltage feedthrough design, the installation of six magnets in long vacuum cavity design is discussed.

## 1 Introduction

The CSNS/RCS is composed of Linac, RCS and target station. The components layout about extraction of the RCS which distribute in one of accumulator ring's straight sections is shown in the Fig.1. It is a total of eight kicker magnet modules to kick the circulating beam vertically such an angle into the downstream extraction septum (Lambertson magnet). Then the septum will deflect the beam horizontally an angle of 15 degrees and make the beam leave the RCS completely [1].

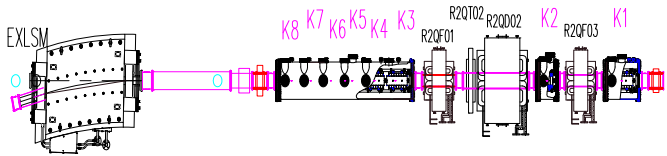

Fig.1. The components layout of the extraction.

The extraction kickers will work on 25 Hz rate with a flat top of 600 ns and a rise time of 265 ns. So the ferrite that has high frequency response and little loss is used as the core material. The eight kicker magnets are installed inside three vacuum tanks that connect with vacuum chamber, but not put the vacuum chamber into kicker magnet. Because the extraction kickers are high frequency AC magnets, so if place vacuum chamber into the magnet, that must use ceramic vacuum chamber structure, which will bring a series of problems such as the bigger magnet aperture, a much higher voltage, more expensive cost and so on. The high voltage feedthrough with three electrode structure is designed to connect kicker magnet with high voltage power supply reliably.

## 2 Kicker specification

Considering beta-function and acceptance about the beam, the 8 kicker magnets have various apertures, effective lengths and relative heights to the beam. To simplify the kicker design, the 8 kicker magnets are grouped in 5 different types[2]. The parameters of the kicker assembly with a typical magnet K1 are listed in Table 1. Each magnet needs a separate power supply [3].

Table 1. The required specifications of the kickers

| Max. field | 558 G |
| --- | --- |
| Effective Length of Magnet | 0.52 m |
| Peak Current | 6660 A |
| Peak Voltage | 45 kV |
| Magnet width | 143 mm |
| Magnet gap | 155 mm |
| Turns per coil | 1 |
| Rise Time of Field | <265 ns |
| Top Time of Field | >600 ns |
| Field uniformity(60% width) | ±1% |
| Repetition rate | 25 Hz |

___________________________

1) E-mail: wanglei@ihep.ac.cn

The magnet inductance is suggested to be measured

$$L = N^2 \mu_0 \omega l_c / h \qquad (1)$$

Where $\mu_0$ is the initial permeability, $\omega$ is the magnet width, $h$ is the magnet gap and $l_c$ is the effective length of magnet.

## 3 Design of the kicker magnet

The fast kicker magnet is designed as rectangle frame magnet with a ferrite core [4]. The cross section and 3 dimensional structure model of kicker magnet are shown in Fig.2.

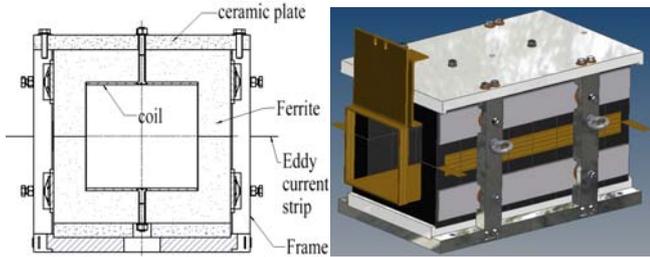

Fig.2. The structure of kicker magnet

The Ni-Zn type ferrite is used for this magnet, which was successfully developed after a lot of tests. This type ferrite provides little outgassing rate, high frequency response and high resistivity, which are important for fast pulsed magnet in high vacuum environment and actual application. The inner surfaces will be coated with TiN films to reduce secondary electron yield. Along the direction of beam, ferrites assembly compose kicker magnets with different effective length and aperture. But the process for sintering a C- shaped ferrite block is very difficult, because the block size is too large. Moreover, the block thickness has a limit value, because thick ferrite block easily generates cracks inside it when sintered. In one experiment after another, we found that the block thickness maximum size of C- shaped ferrite should be smaller than 55 mm. Early drying treatment of the ferrite block is also very important.

In order to install the coil, magnet core can be split into upper and lower halves. In the middle of the two C- shaped ferrite block, two 0.5mm copper strips that connect with vacuum tanks at last are used to carry the beam-induced image current and also to reduce the coupling between the ferrite and the beam. Ceramic plates are added in the magnet above and below, which are used for fixing on coil and electric insulation. The coil is made of a single–turn copper conductor. Cooling coil is not needed because its power is very small.

The simulation results with OPERA-3D [5] about K1 are shown in Fig.3. If the core length is set to 460 mm, magnet width is set to 143 mm, magnet gap is set to 155 mm and powered by 6600 A current, the result after simulation shows that effective magnetic length is 520 mm, and the center field of the magnet is 558 G. Maximum field intensity around the ferrite is 2584 G, that is located in its 4 corners. But the saturation magnetic flux ferrite is 3000 G which has been obtained by measurement and experiment. Field uniformity for 60% width after many times simulation is ±0.7%. So the result of simulation shows that its scheme design is to meet the physical requirements.

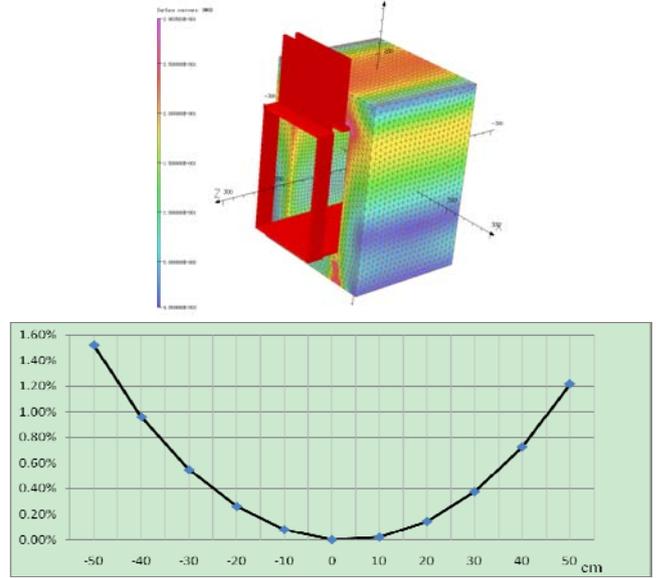

Fig.3. The results of OPERA-3D simulation about K1

The eight kicker magnets are installed inside three vacuum tanks that connect with vacuum chamber. The first two magnets are placed each into a vacuum tank, and other magnets are placed in the third vacuum tank that has about 3 meters long (Fig.4). Before insertion, all the magnets are placed on one long plate. According to their final position, each magnet is aligned and fixed on the plate with adjusting device. Then the plate, with the magnet on it, is horizontally pushed into the tank by the sliding track. The sliding track is composed of a sliding bolt assembly and an adjusting bolt assembly (Fig.5). The sliding bolt assembly is used for smooth pushing the plate with the magnet on it. The adjusting bolt assembly which is collimated at first is used to carry and adjust the plate. Then at last, the plate is fixed reliably.

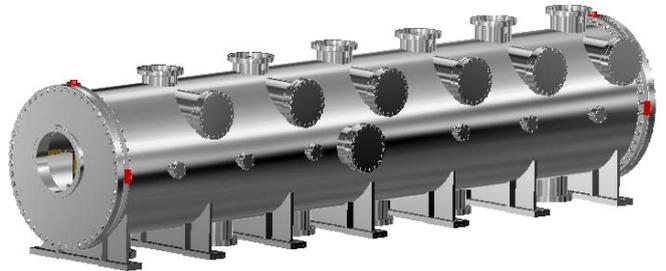

Fig.4. The 3 meters long vacuum tank

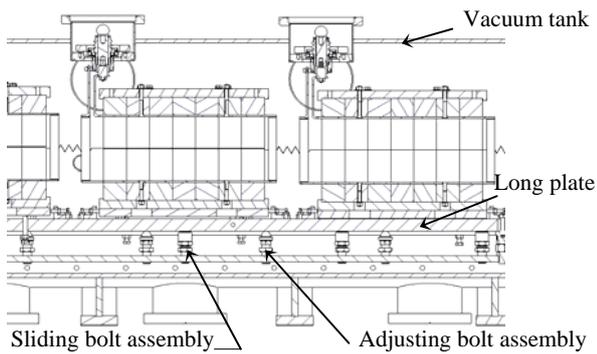

Fig.5. Construction about the sliding track

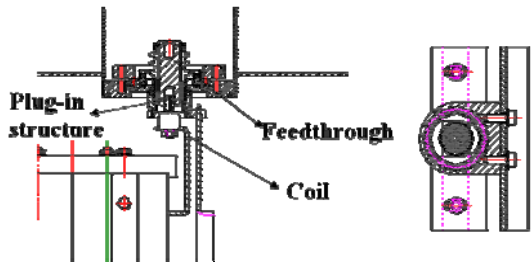

Fig.6. The plug-in structure

Lead wire connection for feedthrough and the magnet coil uses a coaxial plug-in structure without any bolts and nuts. The inner conductor is connected with elastic contact part that fixed on the magnet coil, the outer conductor is connected with hollow metal braid embedded in another part that fixed on the magnet coil. This structure is not only very compact but also easy to install and disassemble (Fig.6).

The kicker magnet is powered by a pulsed power supply. The 45 kV pulsed voltage coming from the power supply will be transmitted into the magnet that is located in the vacuum tank by the feedthrough. The feedthrough was designed with a tri-polar structure, which isolate the ground of the pulsed power supply from the vacuum tank. Instead of the traditional use of precious metals such as molybdenum and titanium, the inner and outer conductors in the feedthrough are made of oxygen-free copper that cost very little and are convenient for machining. Fixing and insulating material between them is ceramics. However, the expansion coefficient of copper and ceramic is not same, so it is difficult to solder them together. Oxygen free copper thickness should be as thin as possible without losing its rigidity. The ceramics that be welded with copper conductor adopt a symmetrical structure that is two ceramic pieces with a thin copper in the middle. That greatly improve the success rate of the welding. So after giving full consideration to this point in design, it has been successfully welded one. The developed high-voltage feedthrough is shown in Fig.7. In addition, the inductance is very small, only 0.15 microhenry. Inside and outside conductor of feedthrough can withstand 25kV after many high-voltage experiments.

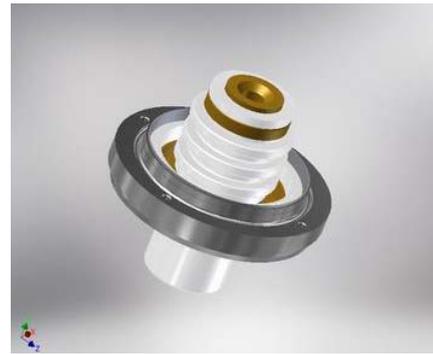

Fig.7. The developed high-voltage feedthrough

Six magnets fixed on a long plate need to be lifted and pushed into the third vacuum tank. Because the plate has three meters long, a suitable lifting point has a great relationship to the overall deformation of the plate. Through simulation analysis, the maximum deflection is 0.2mm at the optimal two points for lifting the long plate (Fig.8). The optimal two points is located at 750mm and 2250mm along the plate. However, if the plate is lifted from both ends, then the maximum deflection will be increased to 2.6mm, that will be dangerous for all the magnets.

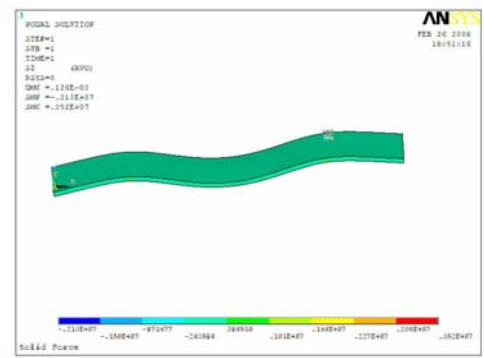

Fig.8. Stress analysis about the long plate

## 4 Vacuum quality criteria

The required vacuum quality for extraction kicker is in the $1\times10^{-6}$Pa range. But large quantities of ferrite and also ceramic plates are used in the tank. A total of eight ion pumps are installed outside of three assemblies to overcome the outgassing from these materials. The vacuum tanks and flanges are made of 304L stainless steel. And flanges are wire seal type with copper gaskets. All blind holes in the tanks are drilled with venting paths to preventing the generation of air entrainment. Weldment about the vacuum tank will follow a principle that is an internal continuous welding and an external spot welding. Ferrite will be cleaned and baked to 120°C slowly before installation. After sealing, all the assemblies can be baked to 180°C in the ring.

## 5 Conclusion

Extraction kicker magnets are very important in CSNS/RCS. Its performance is directly related to the smooth operation of the accelerator. Eight kicker magnets with 5 different types are used to realize the fast extraction out of RCS. Physical and structural designs of eight kicker magnets that located in three vacuum tanks are presented and analyzed. The sliding track composed of a sliding bolt assembly and an adjusting bolt assembly is introduced that is used for smooth pushing in the plate with the magnet on it. Eight kicker magnets are produced now, that will be completed in the near future. There is always work to be done.